# A New Model of Urban Population Density Indicating Latent Fractal Structure

Yanguang Chen

(Department of Geography, College of Urban and Environmental Sciences, Peking University. Beijing 100871, China. Email: chenyg@pku.edu.cn)

**Abstract**: Fractal structure of a system suggests the optimal way in which parts arranged or put together to form a whole. The ideas from fractals have a potential application to the researches on urban sustainable development. To characterize fractal cities, we need the measure of fractional dimension. However, if the fractal organization is concealed in the complex spatial distributions of geographical phenomena, the common methods of evaluating fractal parameter will be disabled. In this article, a new model is proposed to describe urban density and estimate fractal dimension of urban form. If urban density takes on quasi-fractal pattern or the self-similar pattern is hidden in the negative exponential distribution, the generalized gamma function may be employed to model the urban landscape and estimate its latent fractal dimension. As a case study, the method is applied to the city of Hangzhou, China. The results show that urban form evolves from simple to complex structure with time.

## 1 Introduction

Fractal indicates the optimum structure in nature, and a fractal body can fill space in the most efficient way (Chen, 2008a; Rigon *et al*, 1998). The concepts from fractal can be employed to optimize the spatial structure of cities in future city planning. In this sense, fractal theory is helpful for us to study urban sustainable development. Cities have been demonstrated empirically to be of self-similarity (e.g. Batty, 2005; Batty and Longley 1994; Chen, 2008a; Frankhauser 1994). For the fractal bodies, the conventional measures such as length, area, size, and density are always disabled. In this case, fractal dimension is a valid parameter to characterize urban growth and form. In many cases, we can approximately calculate fractal parameters by using the common methods including the *grid method* and *radial method* (Frankhauser, 1998). However, if fractal structure of city agglomerations are concealed by random noise interference or complex spatial distributions of geographical phenomena, the conventional methods will be helpless, we need other approaches such as spectral analysis to estimate fractal dimension indirectly (Chen, 2008b).

To research fractal structure of city systems, we have to build mathematical models. For human geographical phenomena, the models are not one and only. We have more than one equation to model the size distribution of cities, and we have many a function to formulate urban population density (e.g. Carroll, 1982; Batty and Longley, 1994; Cadwallader, 1997; Gabaix and Ioannides, 2004; Zielinski, 1979). Moreover, the parameter values of the model are not real constants. The diversity of models and variability of model parameters suggest the complexity of human geographical systems. Fractal geometry is a powerful tool for us to explore spatial complexity of cities. Fractal dimension is the basic and important parameter to characterize urban fractal structure. The precondition of estimating the fractal dimension of a city is that it follows a power law indicative of some scaling relation between the conventional measures (e.g. length, area, density).

One of the foundations of urban structure studies is to model urban form. Population and land use densities are two central measures for urban form modeling. The land use patterns generally follow the power law and take on clear fractal properties (e.g. Batty and Longley 1994; Benguigui *et al*, 2000; De Keersmaecker et al, 2003; Frankhauser, 1994; Thomas *et al*, 2007; Thomas *et al*, 2008). However, urban population distribution seems to be more complicated. Modeling urban population density is an old question. There are varied opinions on model selection of urban density (Batty and Kim, 1992; Cadwallader, 1997). Following the negative exponential model on urban density distribution (Clark, 1951), a number of mathematical models as revisions or variants to Clark's law are proposed. The noticeable models include the Gaussian formulation (Sherratt 1960; Tanner 1961), the inverse power function (Smeed, 1961; Smeed, 1963), and the quadratic exponential equation (Latham and Yeates 1970; Newling 1969). The more general expression may be the gamma model, which is used to reconcile the debates between the negative exponential and inverse power-law distributions (Batty and Longley, 1994; Tanner, 1961).

An interesting discovery is that the gamma model can be expanded and then used to estimate the latent fractal dimension of urban form indirectly. In the paper, a possible model is proposed to characterize the urban density, settling the arguments between several urban population density models. The study starts with the negative exponential function. The Clark's model is generalized to a negative exponential model with power by combining it with the Sherratt's model. Then, through taking the Batty-Longley's model on fractal urban form as a weight function, the exponential-power model is further developed to a new model, which can be employed to estimate the fractal dimension of urban form and predict the spatial complication of urban evolvement. As an empirical analysis, the new model is applied to the city of Hangzhou in China. The results can bring us a better understanding of the basic principles on urban population density as well as the urban development patterns.

## 2 Mathematical models

### 2.1 Generalization of the negative exponential model

Among all the models characterizing the functional relationship between population density and distance, the Clark's law bears both elegant form and theoretical foundations. The negative exponential function can be derived either from the theories of strict utility-maximizing associated with urban economic theory (Beckmann 1969; Mills 1972; Muth, 1969) or from the operational urban model based on entropy-maximizing (Bussiere and Snickars, 1970; Chen, 2008b). These derivations might constitute the foundation of the clear demonstration that disaggregate models of individual resource allocation in space based on utility-maximizing were consistent with the macro-models of spatial interaction based on ideas from entropy-maximizing (Batty 2000).

However, despite these achievements, new sharp questions were posed such as reconciling the Clark model with the law of allometric growth on the relationships between area and population of urban systems, especially when cities take on fractal patterns. Now that the density of urban land-use can be defined by an inverse power function, a contradiction will emerge immediately between the Clark model and the power law of urban area-population allometry. Batty and Kim (1992) argued that many previous estimates of population density functions should be reworked with power functions. This is an interesting and developmental viewpoint. Besides this, perhaps we should try to find other approaches to solving the problems.

As a theoretical study, for simplicity, this work is only focused on the monocentric cities. The polycentric cities will be studied in a special paper. Assuming that population density $\rho(r)$ at distance $r$ from the center of city (where $r=0$) declines monotonically according to certain relative rate of change, Clark (1951) proposed an empirical expression on urban density as follows

$$\rho(r) = \rho_0 \exp(-br), \qquad (1)$$

where $\rho_0$ is a constant of proportionality, which is expected to equal the central density of a city, i.e., $\rho_0 = \rho(0)$, and the *rate parameter*, $b$, denotes the density gradient, indicating a rate at which the effect of distance attenuates. If we define a characteristic radius as $r_0=1/b$, then the Clark model can be rewritten in the form

$$\rho(r) = \rho_0 \exp(-\frac{r}{r_0}), \qquad (2)$$

which can be derived by an entropy maximizing method (Chen, 2008b). It is easy to prove that the negative exponential function is equivalent to a spatial autocorrelation function, and the *scale parameter*, $r_0$, can be associated with the spatial correlation length (Chen and Zhou, 2008).

Many negative exponential models can be derived from the entropy maximizing hypothesis. From the view of cities, the distance variable can be defined between 0 and infinity ($0 \leq r < \infty$). Thus, the Clark's model can be derived by an entropy maximizing method. However, from the view of regions, the distance measure can be defined between negative infinity and positive infinity ($-\infty < r < \infty$). In this instance, we can derive Sherratt's model by using the principle of entropy maximization. The Sherratt model is in fact a Gaussian function (Sherratt, 1960; Tanner, 1961)

$$\rho(r) = \rho_0 \exp(-\frac{r^2}{2r_0^2}), \qquad (3)$$

in which the constants $\rho_0$ and $r_0$ fulfill the same roles as in equation (2). Apparently, the Clark model and the Sherratt model share the similar mathematical expression.

If we regard entropy-maximization as a cause or a process rather than a real state or a final result, we had better attach a constraint parameter to the variable of equation (2). The parameter value varies from 1 to 2 in theory. Consequently, both the Clark's model and the Sherratt's model are extended to more general form. Actually, because of fractality of urban form, the scale parameter in equation (2) is not real constant, but a variable depending on the urban radius we determined. In other words, there is indirect scaling relation between $b$ and $r$. The relation can be revealed as

$$b = kr^{-v}, \qquad (4)$$

where $k$ and $v$ are two parameters. Substituting equation (4) into equation (1) yields (Chen, 1999)

$$\rho(r) = \rho_0 \exp(-kr^{1-v}) = \rho_0 \exp(-\frac{r^\sigma}{\sigma r_0^\sigma}), \qquad (5)$$

where $\sigma=1-v$ is the constraint parameter, $r_0=(\sigma k)^{-1/\sigma}$ is the rescaled characteristic radius (a new scale parameter). Of course, the coefficient $\rho_0$ is still the density at the origin since the parameter $\sigma$ is greater than zero, i.e. $\sigma>0$. Equation (5) can be called the exponential-power model, which shows that the population density of a city is given by the exponential decay function of distance to the power $\sigma$. Apparently, if $\sigma=1$, equation (5) will reduce to Clark's model, and if $\sigma=2$ as given, then equation (5) will become Sherratt's model. It is expected that $\sigma$ comes empirically between 0 and 2. Clark's model can be applied to many cities in the real world, while the Sherratt model has the advantage of a simpler expression for mathematical analysis (Dacey, 1970).

## 2.2 A new model for urban fractal dimension estimation

Urban population depends on urban land for existence and *vice versa*. There is an interaction between human beings and land use of cities. Taking real urban form into consideration, we should introduce Batty-Longley's models of urban land-use patterns into the urban density model. The area-radius scaling relation of fractal cities can be written as (Batty and Longley 1994; Frankhauser 1998; Longley *et al*, 1991)

$$N(r) = Kr^{D_f}, \qquad (6)$$

in which $r$ refers to radius, $N(r)$ to the actual land-use area within the circle of radius $r$, and the area can be represented by the number of pixels in digital map, $K$ denotes a proportionality coefficient, and $D_f$ the parameter of fractal dimension which scales land-use area with distance. Generally speaking, the fractal dimension is less than 2 but greater than 1, namely $1<D_f<2$.

However, as a *radial dimension* (Frankhauser, 1994), the fractal parameter can exceed the upper limit 2 in some special cases (White and Engelen, 1994). According as equation (6), the land-use density $\rho_L$ can be given by the derivative of $N(r)$ with respect to area unit $(dr)^d$, and we have

$$\rho_L(r) = \frac{dN(r)}{d^d r} \propto r^{D_f - d}. \tag{7}$$

where $\rho_L(r)$ is the land-use density at distance $r$ from the center of city ($r=0$), $d=2$ denotes the Euclidian dimension of the embedding space of urban form. However, there is no definition for the central density of a city in the inverse power function. The central density should be defined separately.

Suppose that urban population is distributed on the fractal land-use patterns. The form of land use is determined by population distribution and in turn reacts on it. Thus population density function can be used as the weight of land use density distribution, and land use density function can also be used as the weight of population density distribution. That is, there exists a relationship of "mutual weight" between urban population and land use density. Combining equation (7) with equation (5) yields a weighted exponential-power model

$$\rho(r) = C r^{D_f - d} \exp(-\frac{r^\sigma}{\sigma r_0^\sigma}), \tag{8}$$

where $C$ is the proportionality coefficient, other parameters play the same parts as in equation (5). This model can be regarded as "**generalized gamma function** (GGF)" that is general enough to encompass the various arguments about one functional form or the other except Newling's model. The parameter $D_f$ is in fact fractal dimension hidden behind quasi-exponential distribution, so it can be termed as *latent dimension* of urban form.

The GGF differs from the standard gamma function because of the constraint parameter $\sigma$, but the basic characters of the gamma function remain (Figure 1). If $\sigma=1$, equation (8) will be reduced to the common gamma function, which is actually a special spatial correlation function. The new model fails to define the point where $r=0$, or we should define $\rho(0)=\rho_0$ as a complement. If the fractal dimension of urban land-use reaches the Euclidean dimension, i.e. $D_f \to d=2$, the model will collapse to equation (5); but if the characteristic radius of urban population distribution approaches infinity, namely $r_0 \to \infty$, then the model will evolve into Smeed's (1963) formulation such as

$$\rho(r) = C r^{-\alpha} = C r^{D_f - d}, \tag{9}$$

where $C$ is the proportionality coefficient, the parameter $\alpha=d-D_f$ refers to the scaling exponent. GGF can contain at least four urban density models with which geographers are familiar (Table 1). The variety of the urban population density model and the variability of the model parameters suggest asymmetry of human geographical systems, and asymmetry or symmetry breaking indicates spatial complexity and complication (Chen, 2008a).

Table 1 Urban density models encompassed by the new model (GGF)

| Parameter | Function | Originator, Pioneer or Urger |
| --- | --- | --- |
| $D_f = d$, $\sigma = 1$ | $\rho(r) = \rho_0 \exp(-r/r_0)$ | Clark (1951) |
| $D_f = d$, $\sigma = 2$ | $\rho(r) = \rho_0 \exp[-r^2/(2r_0^2)]$ | Sherratt (1960); Tanner (1961); Dacey (1970) |
| $D_f < d$, $\sigma = 1$, $r_0 \to \infty$ | $\rho(r) = Cr^{D_f - d}$ | Smeed (1961, 1963); Batty and Longley (1994); Frankhauser (1998) |
| $D_f < d$, $\sigma = 1$ | $\rho(r) = Cr^{D_f - d} \exp(-r/r_0)$ | Tanner (1961); March (1971); Angel and Hyman (1976) |
| $D_f = d$, $0 \le \sigma \le 2$ | $\rho(r) = \rho_0 \exp[-r^\sigma/(\sigma r_0^\sigma)]$ | Chen (1999), Feng (2002) |
| $1 < D_f < 3$, $\sigma = 1$ | $\rho(r) = Cr^{D_f - d} \exp[-r^\sigma/(\sigma r_0^\sigma)]$ | Author of this article |

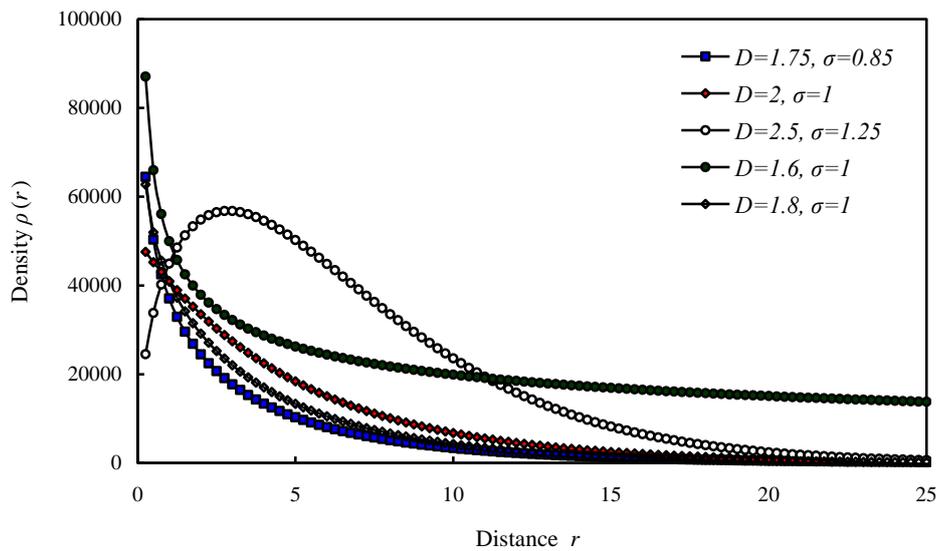

Figure 1 The patterns of the GGM with different parameter values ($\rho_0$=50000 persons/km$^2$, $r_0$=5 km)

**Note**: The scale parameter is taken as $r_0$=5 for all the lines but the fourth one with $r_0$=500000. The fractal dimension $D_f$ is based on the area-radius scaling. It is different from the box dimension in value. The value of the radial dimension can be greater than 2 if the measuring center is not the centroid of urban area.

The new model can be used to reconcile the debates between the negative exponential distribution and the inverse power distribution of urban density. Though a great majority of cities in the real world seem to be in favor of Clark's model, there once appeared other opinions. Parr (1985) argued that the population density in the urban area itself prefers to confirm the negative exponential function, while that of urban fringe and hinterland tends to obey the inverse power law. Another interesting viewpoint was proposed by Longley and Mesev (2000), who claimed that the density function has two parts: for small $r$, $\rho(r)$ decreases as Smeed's model without time-dependence, while for larger $r$, $\rho(r)$ decreases more swiftly (see also Benguigui *et al*, 2001).

The author's viewpoint is that the negative exponential model and the inverse power model represent different spatial modes and states. In fact, a spatial process may not be stationary, though stable (Haining 1990); and city fractals are evolving phenomena through self-organizing process (Benguigui, *et al*, 2000). If the characteristic radius $r_0$ in equation (8) gets larger and larger, the spatial distribution of urban population has a tendency of evolving into a self-similar state. It has been demonstrated that the density gradients of cities become smaller and smaller (Berry, *et al*, 1963; Bank, 1994). This implies that, as expected, the characteristic radiuses do become longer and longer until the generalized gamma distribution turns to the inverse power-law distribution.

A significant use of GGF is that it can be employed to estimate indirectly the fractal dimension of urban form. The new model can play its part of evaluating fractal dimension in three cases. First, there are density data of urban population but there is an absence of land use data. In this instance, if urban population density follows the negative exponential distribution approximately, we can use GGF to estimate the fractal dimension of land use form. Second, the fractal pattern of urban form (say, land use density) is concealed by negative exponential phenomena (say, population density). In this case, we had to use generalized gamma model (GGM) to estimate fractal dimension. Third, urban structure takes on self-affinity, which can be associated with exponential distribution or quasi-exponential distribution. For the purpose of verifying the above conjecture, GGF will be applied to Hangzhou, the capital of Zhejiang Province, China.

## 3 Empirical Analyses

### 3.1 Study area and methods of data processing

The city of Hangzhou is chosen for our empirical analyses because its population density data has already been fitted to the Clark's model (Feng, 2002). Four sets of census data of the city in 1964, 1982, 1990, and 2000 are available. The census enumeration data is based on *jie-dao*, or sub-district (Wang and Zhou, 1999), which bears an analogy with urban zones in Western literature (Batty and Longley, 1994). In fact, a zone or sub-district (*jie-dao*) is an administrative unit comprising several city blocks defined by streets and other physical features. The study area is confined in the combination of city proper and its outskirts, and this scope comes approximately between the urbanized area (UA) and the metropolitan area (MA) of Hangzhou. The zone with maximum population density, which is very close to the urban functional core, is defined as the center of the city, and the data are processed by means of spatial weighed average. The method of data processing is illuminated in detail by Feng (2002). The length of sample path is 26, and the maximum urban radius is 15.3 kilometers. The basic set of zones for population data are shown in figure 2, which displays aggregation of census data.

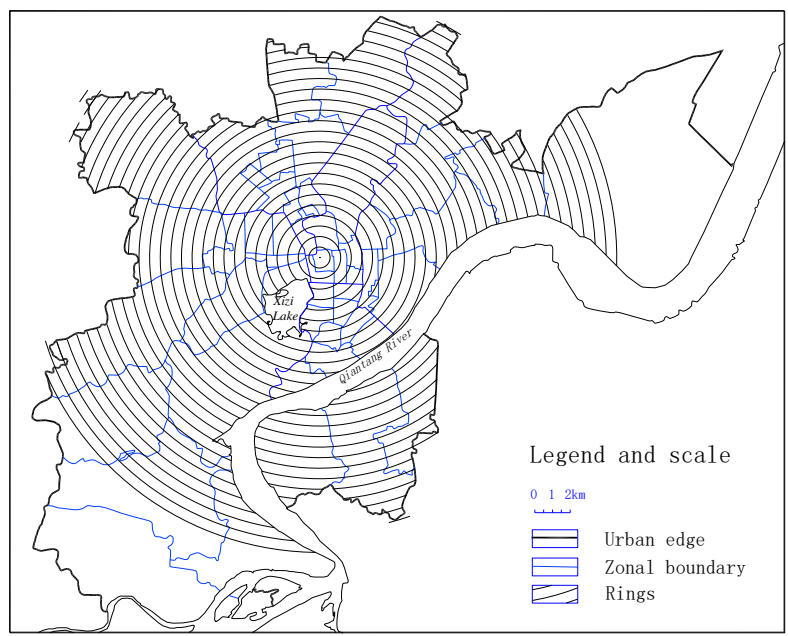

Figure 2 The choropleth census map of the Hangzhou metropolis (by Feng, 2002)

For simplicity, we can number the rings as $i$ ($i = 0,1,2,\cdots,25$, where $i$=0 implies the center of circles), and number the zones as $j$ ($j = 1,2,\cdots,47$). Given

$$S_{ij} = R_i \cap Z_j, \qquad (10)$$

where $R_i$ refers to the $i^{th}$ ring, and $Z_j$ to the $j^{th}$ zone measured with area, it follows that $S_{ij}$ represents the intersection of $R_i$ and $Z_j$. For the $i^{th}$ ring, defining a weight as

$$w_{ij} = \frac{S_{ij}}{\sum_j S_{ij}}, \qquad (11)$$

we have

$$\bar{\rho}_i = \sum_j w_{ij} \tilde{\rho}_j, \qquad (12)$$

where $\tilde{\rho}_j$ represents the population density of the $j^{th}$ zone, and $\bar{\rho}_i$ denotes the weighted average density of the $i^{th}$ ring. The computations form four urban population density samples (see Appendix 1).

### 3.2 Computations and analysis

We have two approaches for fitting the urban density data to the models, including the negative exponential and the exponential-power model. One approach is the curvilinear regression based on the least square method, the other is non-linear curve fit based on iteration technique. For a great majority of cases, the two approaches are not equivalent, thus the results are different. The choice of the specific method is based on study objective. Generally speaking, the former approach is for theoretical explanation, while the latter one is for practical prediction. The ordinary least square (OLS) method will be employed to estimate the values of the model parameters for the purpose of theoretical research.

First, we should decide which model is more appropriate to the population density of Hangzhou city. As experimentation, various functions possible for modeling urban density, including logarithmic function, power function, normal function, and lognormal function, are tested one by one by regression analysis. On the whole, the modeling results are not satisfying, or the physical meaning of the parameter values can not be explained. However, when the data is fitted to the Clark model, the results are acceptable (Feng, 2002). The evaluation of the coefficients $\rho_0$ and $r_0$ of the negative exponential model is relatively simple. Turning the model into the linear form by means of log-transformation, we can perform the linear regression using any software for mathematical or statistical analysis. As for the parameters of exponential-power model and the new model, it is not very convenient. Let $k=1/(\sigma r_0^\sigma)$, then the characteristic radius $r_0$ in equations (5) and (8) can be given by $r_0 = (\sigma k)^{-1/\sigma}$.

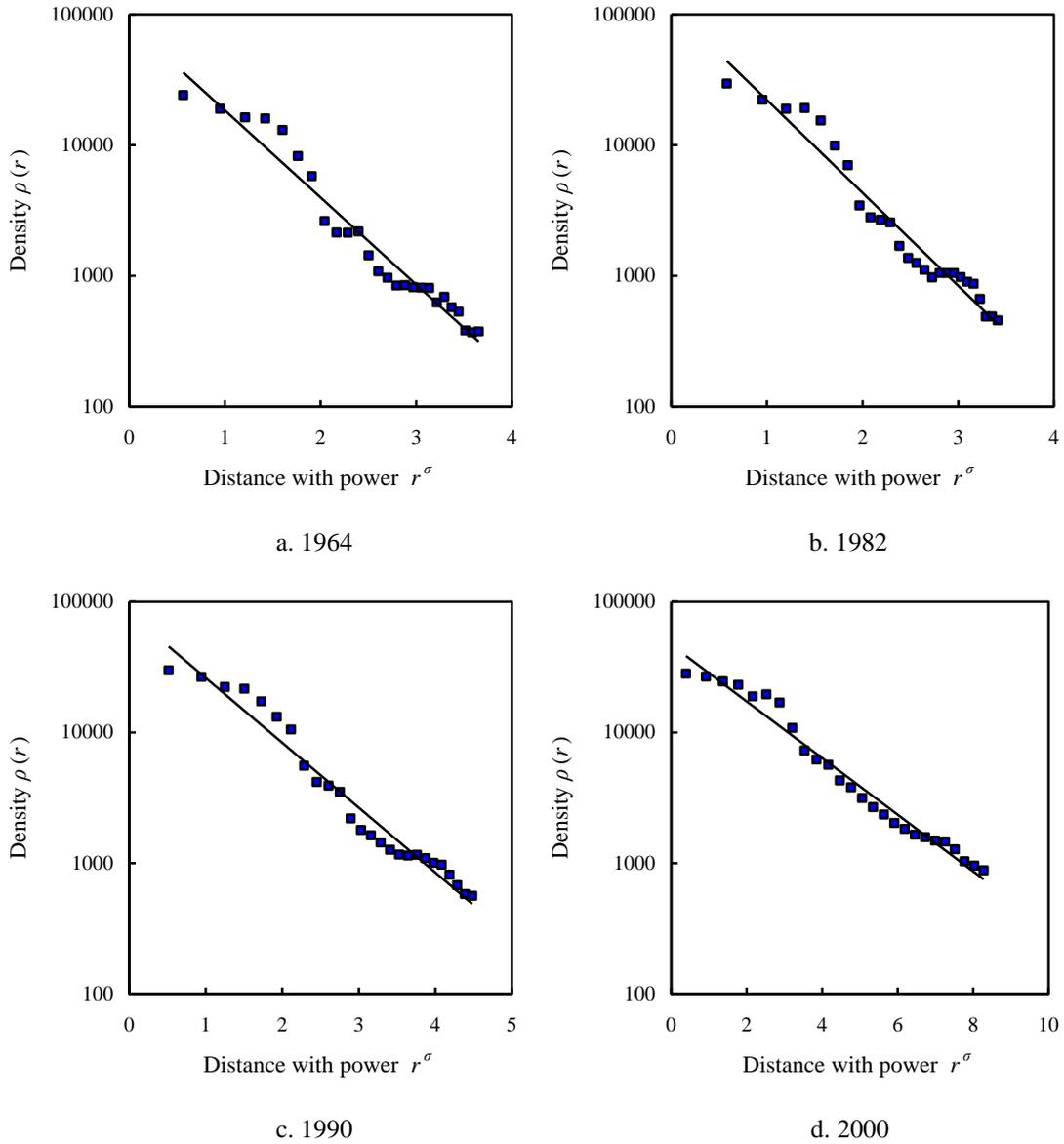

a. 1964

b. 1982

c. 1990

d. 2000

**Figure 3 The semi-logarithmic plots based on the exponential-power model of urban population density for Hangzhou in four years**

Fitting the urban density data to equation (2) and equation (5) respectively by the least square calculation yields the estimated values of the parameters in the models (Table 2). For visual examination, the data points can be displayed on two kinds of plots: arithmetic plot and semi-logarithmic plot. In the semi-logarithmic plot, the negative exponential function seems not to be the best selection for the urban density of Hangzhou (Appendix 1), but the goodness of fit goes up as a whole from 1964 to 2000 (Table 2). If we use the exponential-power model, i.e. equation (5), to replace the negative exponential model, equation (2), we can get better effect of fit, which is shown in figure 3(Appendix 2). However, the criterions for model selection cannot be confined to statistics such as goodness of fit, standard error, etc. As Samuel Karlin (1983) pointed out: "The purpose of models is not to fit the data, but to sharpen the questions." (Quoted from Buchanan, 2000, page 85)

**Table 2 Estimated values of model parameters for urban density of Hangzhou city in four years**

| Models | Parameters and statistics | 1964 | 1982 | 1990 | 2000 |
|---|---|---|---|---|---|
| Negative exponential model | Proportionality constant $\rho_0$ | 16430.945 | 19493.134 | 24876.076 | 31849.983 |
| | Characteristic radius $r_0$ | 3.565 | 3.671 | 3.628 | 3.946 |
| | Determination coefficient $R^2$ | 0.907 | 0.904 | 0.930 | 0.968 |
| Exponential-power model | Proportionality constant $\rho_0$ | 85633.458 | 113984.742 | 82025.360 | 46867.860 |
| | Characteristic radius $r_0$ | 1.946 | 1.971 | 2.326 | 3.409 |
| | Constraint parameter $\sigma$ | 0.475 | 0.450 | 0.550 | 0.775 |
| | Determination coefficient $R^2$ | 0.958 | 0.956 | 0.965 | 0.976 |

To calculate the fractal dimension of urban land-use structure through the weighed exponential-power model, GGF, we need at least two statistical criteria: one is the global statistic--determination coefficient, $R^2$, which must pass the statistical test; the other is the local statistic--$P$-values of the regression coefficients indicative of significance, which must be less than 0.05. The criterions can be summed up as "Maximize $R^2$ subject to $P$-value <0.05", according to which we determine the values of the constraint parameter $\sigma$. Taking natural logarithms on both sides of equation (8) gives a two-variable quasi-linear regression equation

$$\ln \rho(r) = \ln C + (D_f - d)\ln r - \frac{r^\sigma}{\sigma r_0^\sigma}, \tag{13}$$

by which the latent fractal dimension of urban form, $D_f$, can be estimated. As for Hangzhou, a least squares computation gives the following results, which is listed in Table 3. The trend lines on the semi-logarithmic plots are not straight due to nonlinear relationships (Figure 4). The results suggest that the urban land use has been spreading (the fractal dimension $D_f$ becomes larger and larger) while the population has been concentrating (the characteristic radius $r_0$ becomes smaller) despite some evidence for suburbanization round about 2000 (Feng, 2002; Feng and Zhou, 2005).

**Table 3 Estimated values of fractal dimension and related parameters of the new model of Hangzhou's urban form**

| Parameter name | 1964 | | 1982 | | 1990 | | 2000 | |
|---|---|---|---|---|---|---|---|---|
| | Value | $P$-value | Value | $P$-value | Value | $P$-value | Value | $P$-value |
| $C$ | 19836.537 | 0 | 23523.569 | 0 | 28616.853 | 0 | 30618.166 | 0 |
| $r_0^*$ | 6.131 | 0 | 6.444 | 0 | 5.312 | 0 | 4.872 | 0 |
| $D_f$ | 1.373 | 0.001 | 1.374 | 0 | 1.533 | 0.003 | 1.784 | 0.040 |
| $\sigma$ | 1.0 | | 1.0 | | 1.0 | | 1.1 | |
| $R^2$ | 0.945 | | 0.944 | | 0.953 | | 0.968 | |

**Note**: The symbol $r_0^*$ denote the rescaled characteristic radius based on GGF.

To analyze the spatio-temporal evolution of Hangzhou's urban form, we can collect the estimated results of the principal parameters and tabulated them as follows (Table 4). Further, we can calculate the *information entropy* of Hangzhou's population distribution by using the following formula

$$H = -\sum_{i=0}^{n} p_i \log_2 p_i, \qquad (14)$$

where $H$ denotes information entropy, $i=0,1,\ldots,n$ (the number of rings is $n+1=26$), and the "probability" $p_i$ is defined by

$$p_i = \bar{\rho}_i / \sum_{i=0}^{n} \bar{\rho}_i. \qquad (15)$$

in which the symbol $\bar{\rho}_i$ fulfills the same roles as in equation (12). The results are also displayed in Table 4 for comparison with fractal dimension.

**Table 4 Characteristic radius, latent fractal dimension, and information entropy of Hangzhou, 1964-2000**

| Year | Characteristic radius ($r_0$) | Rescaled characteristic radius ($r_0^*$) | Latent fractal dimension ($D_f$) | Information entropy ($H$) |
|---|---|---|---|---|
| 1964 | 5.941 | 6.131 | 1.373 | 3.548 |
| 1982 | 6.119 | 6.444 | 1.374 | 3.584 |
| 1990 | 6.048 | 5.312 | 1.533 | 3.677 |
| 2000 | 6.576 | 4.393 | 1.862 | 3.931 |

**Note**: The unit of the characteristic radius and the rescaled characteristic radius is "kilometer (*km*)", and the information unit is "bit".

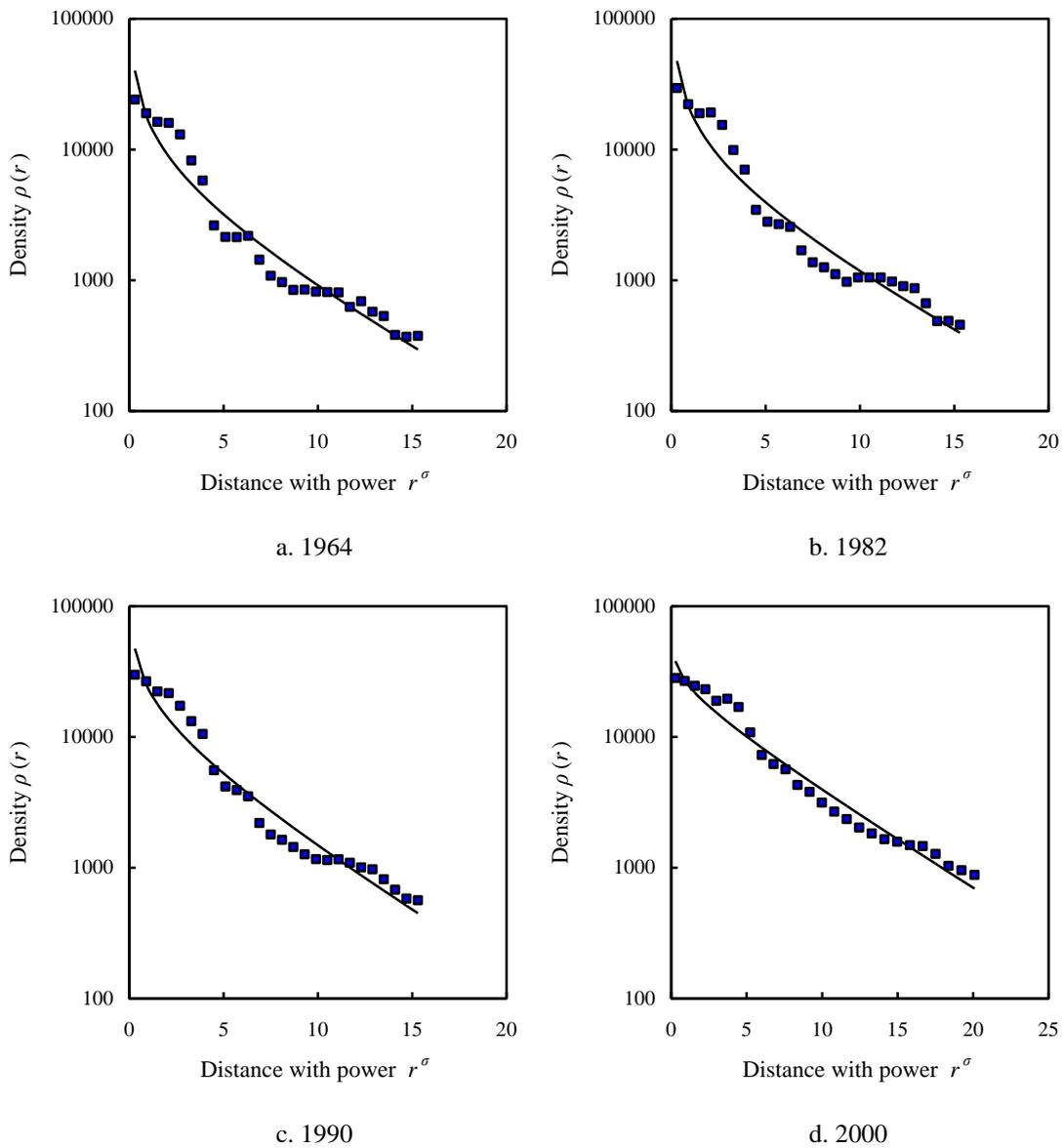

a. 1964    b. 1982

c. 1990    d. 2000

**Figure 4 The semi-logarithmic plots based on GGF of urban density for Hangzhou in four years**

There seems to be some contradiction between the results based on the negative exponential model and those based on GGF. According as the negative exponential model, the characteristic radius became larger as a whole from 1964 to 2000 (Table 2). However, according as the GGM, the rescaled characteristic radius became smaller from 1982 to 2000 (Figure 5a). Generally speaking, the conclusion drawn from the negative exponential model should correspond to that from the new model. In fact, the population in Hangzhou region kept concentrating at large scale from 1975 to 2000 (Chen and Jiang, 2009). In this sense, the result indicates that the characteristic radius becomes smaller for the time being may be more consistent with the reality. On the other hand, the latent dimension went up from 1964 to 2000. The increase trend of the fractal dimension value is linearly correlated with that of the information entropy (Figure 5b). This lends further empirical support to the theoretical relation between information entropy and fractal dimension (Ryabko, 1986). Fractal dimension increased and went towards $d=2$, meanwhile the rescaled characteristic radius descended. All these seem to suggest the trend of Hangzhou's population density evolved from the nonstandard negative exponential distribution into the standard negative exponential distribution within this period. Meanwhile, the information entropy increased, suggesting a process of spatial entropy maximization.

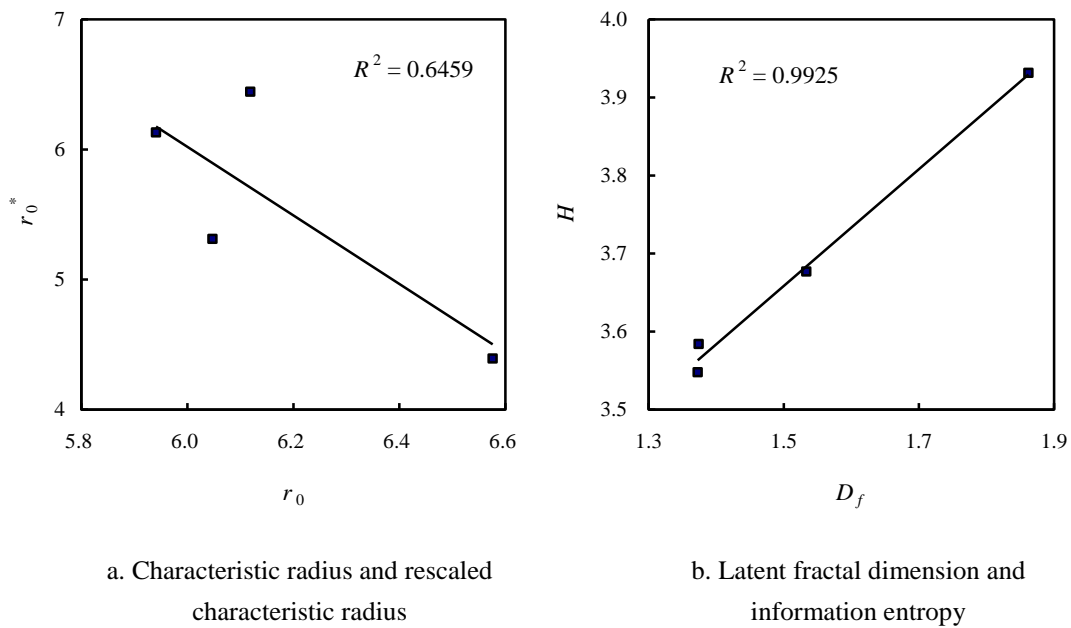

a. Characteristic radius and rescaled characteristic radius

b. Latent fractal dimension and information entropy

**Figure 5 Typical parameter relationships of Hangzhou, 1964-2000**

# 4 Questions and discussion

## 4.1 Entropy maximization, spatial optimization, and 3R cities

The essence of fractals is the scaling symmetry (Mandelbrot and Blumen, 1989). Symmetry is a significant concept for us to understand cities as systems and systems of cities (Chen, 2008a). One of the basic properties of urban evolution is to follow the scaling laws indicative of contraction or dilation symmetry. Symmetry depends on the conservation law, while the universality of natural laws depends on symmetry. Without symmetry there would be no universality of natural laws. Symmetry and symmetry breaking are always related with complexity and complication (Mainzer, 2005, Chen, 2008a). Human geographical systems are of no translational symmetry in both space and time. The translational symmetry seems to be replaced by the scaling symmetry suggesting fractals. Mathematical models of cities are different under different spatial and temporal conditions. Even under the same spatio-temporal condition, an urban phenomenon can be described with different models. The complexity of cities is marked by two aspects: the diversity of mathematical models and the variability of parameter values of a model. Therefore, it is hard to say which model is better for urban population density. Different model has different use and can be employed to characterize different state of the urban phenomenon.

Exponential models indicate the translational symmetry of urban form, and power law relations suggest the dilation/scaling symmetry of urban evolvement (Chen, 2008a). Based on the hierarchical structure of cities, a power law can be decomposed into two exponential laws (Chen and Zhou, 2008). In this sense, translational symmetry lays the foundation of scaling symmetry. A negative exponential model can be derived by an entropy maximizing method (e.g. Bussiere and Snickers, 1970; Chen, 2008b; Curry, 1964; Wilson, 1970). Accordingly, a power law can also be derived from the principle of entropy maximization (Chen, 2009). The spatial optimization of fractal cities can be comprehended from the angle of view of entropy maximization. Note that the entropy of human systems differs from that of physical systems, and entropy maximization of urban systems is dissimilar with the entropy increase in thermodynamics of physical systems (Wilson, 1970; Wilson, 2000). Thermodynamic entropy implies disorder, while human entropy always indicates structure. Let us take the negative exponential model based on Hangzhou city as an example to illustrate this concept.

Suppose that the total population in the urban field of a monocentric city is $P_t$, and the urban growth is considered to be a continuous spatio-temporal process. An urban field is defined as a bounding circle with a radius of $R$ from the city center. In the digital map, we can string $n+1$ pixels indicating cells (Chen, 2008b). Further suppose that the population in the $i$th cell along a radius is $\rho_i$ ($i=0, 1, 2, …, n$), and the whole population on the radius is $P$. Then the *state entropy* of population distribution profile, $H_e$, can be given by

$$H_e = \ln W = \ln P! - \sum_{i=0}^{n} \ln \rho_i!, \tag{16}$$

where $W$ is number of states of the population distributed in all the cells along the radius. Based on equation (16), a nonlinear programming model can be constructed as follows

$$\boxed{\begin{aligned} \max \quad & H_e = P! - \sum_{i=0}^{n} \ln \rho_i! \\ \text{s.t.} \quad & \begin{cases} \sum_{i=0}^{n} \rho_i = P \\ \sum_{i=0}^{n} 2\pi i \rho_i = P_t \end{cases} \end{aligned}} . \tag{17}$$

This denotes that the urban entropy approaches to maximization, subjected to certain total population of the city and the constant average population in every direction. One of the dual forms of the nonlinear programming is

$$\boxed{\begin{aligned} \min \quad & P_t = \sum_{i=0}^{n} 2\pi i \rho_i \\ \text{s.t.} \quad & \begin{cases} \sum_{i=0}^{n} \rho_i = P \\ \ln P! - \sum_{i=0}^{n} \ln \rho_i! = H_e \end{cases} \end{aligned}} . \tag{18}$$

This suggests that the total population within the urban field approaches to minimization, conditioned by certain urban entropy and the determinate population in each direction.

Entropy maximization is the underlying rationale of the Clark's model. Starting either from equation (17) or from equation (18), we can derive equation (1) or (2) (Chen, 2008b). Where mathematical modeling is concerned, equation (17) is equivalent to equation (18). However, as far as physical meaning is concerned, equation (17) is different from equation (18). Derivation of Clark's model from equation (17) suggests that city systems seek equity for individuals (elements). When population size is given, a city tries to maximize its information entropy. Entropy is a measure of conditional uniformity. Entropy maximization indicates the most probable state of urban population distribution. The people living in the central part of a city can enjoy better social service, but have to suffer worse ecological environments. In contrast, the people living in suburbs can enjoy better natural environment but have to suffer worse social service. On the other hand, derivation of the exponential distribution from equation (18) suggests that city systems seek efficiency for the whole. When information entropy is certain, a city tries to minimize its population size, and thus waste least land and resources. This accounts for the fact that a great majority of cities in the world are small urban places rather than metropolises (It is hard to make clear all these questions in a few lines of words, and the discussion will be expanded in a companion paper).

All in all, the principle of entropy maximization is the underlying rationale of city fractals. An exponential model is based on one process of entropy maximization, while a power-law model indicative of fractals is based on two correlative processes of entropy maximization. The state entropy of a city system ($H_e$) defined by equation (16) is proportional to its information entropy ($H$) defined by equation (14), and the information entropy is in the proportion to its fractal dimension (Figure 5). Fractal dimension implies space-filling extent and spatial order. Entropy maximization of cities suggests that a city should possess the minimum population size subject to certain spatial order, or the maximum spatial order subject to certain population size. In light of the essential of entropy maximization of urban evolvement, fractal structure of cities implies the best balance relation between the equity of individuals and the efficiency of the whole of city systems. How to appease the conflict between the equity and efficiency is a difficult problem in economics remained to be solved for a long time. Cultivating fractal structure through self-organized process may be one of the best ways out.

In terms of the above studies, cities can be divided into three categories: real city, regular city, and regressive city. (1) *Real city*: the city in reality. It indicates what a city is really to be at present. A real city may possess well-developed structure, or underdeveloped structure. The Newling's model, or the lognormal model, or one of the models displayed in table 1 can be used to characterize the population density of a real city (table 5). (2) *Regular city*: the well-ordered city with symmetrical structure. It denotes what a city is expected to be according as some theory or model. A regular city bears two characters: first, it can be described by the mathematical models indicating optimized structure, and second, the model parameters such as fractal dimension fall into the proper range. The Clark's model or Smeed's model can be employed to describe the population density of monocentric regular cities. Clark's model is of translational symmetry while Smeed's model is of scaling symmetry. Both these two model are based on entropy maximization, suggesting the good relation between equity and efficiency of urban development. (3) *Regressive city*: a city in transitory state. A real city should evolve around the regular city, or we should urge the real city to evolve towards the regular city. If national planning policies are proper, the real cities evolve close to the regular city. However, if national planning policies are not proper or even wrong, the real cities evolve away from the regular city.

If we plan a real city with the idea from the model of regular city, the real city may come into the state of regressive city. The GGM is suitable to describing the regressive cities coming between the negative exponential distribution and inverse power-law distribution (table 5). For the regressive cities, the model reflecting spatial structure is not very clear. It is often on the line or cuts both ways. In this instance, the fractal structure is usually hidden and the fractal dimension cannot be estimated with the power-law relation.

**Table 5 Three categories of cities and the corresponding models**

| City | Model | State |
| --- | --- | --- |
| Real cities | Clark's model, Sherratt's model, Smeed's model, Newling's model, lognormal model, etc. | Existing state |
| Regressive cities | Exponential-power model, Gamma model, GGM, etc. | Transitory state |
| Regular cities | Clark's model, Smeed's model, etc. | Ideal state or eigen state[*] |

[*]**Note:** The regular cities can be described with an eigenfunction: a negative exponential function or an inverse power function. So the ideal state is termed "eigen state".

## 4.2 GGM indicate state of urban evolvement

The results of modeling Hangzhou's urban form and the change trend of parameters remind us of spatial complexity and complication. The concepts of "complexity" and "complication" come from John von Neumann, who is often considered to be one of the well-known scientists having made the greatest contribution to the topic of complexity in modern science. In von Neumann's work, the terms "complexity" and "complication" are used in the same context and it is difficult to determine whether or not he uses them as synonyms or as two different situations (Israel, 2005). Today we often relate or equal complexity to/with complication: "complex" indicates something consisting of many different and related parts, while "complicate" denotes to make things more difficult or confusing to be understood by making it more complex. However, in recent years, "complication" is endowed with new meaning so that it differs from "complexity" to some extent. Today, the notion of complication suggests the transfer from complex structure to much more complex structure in the evolution of complex systems (Sonis, 2002).

Reconciling the Gaussian function, exponential function, and power function, the new model can be utilized to predict the spatial complication of urban evolution. When the latent dimension $D_f \to d=2$, the population distribution approaches toward a Euclidean plane, and model evolves into the exponential-power model. Then, if the constraint parameter $\sigma \to 2$, the exponential distribution will evolve into the normal distribution. This is a process of degeneration indicating unsustainable urban growth. On the contrary, if the characteristic radius $r_0 \to \infty$, and the fractal dimension comes between 1 and 2, the exponential distribution will evolve into the power-law distribution indicative of fractal landscape. This is a process of advancement and complication indicative of sustainable urban development. In fact, the emergence of fractal structure suggests spatial complexity. Complexity is a relative conception. Compared with the normal distribution, the exponential distribution suggests complexity (Goldenfeld and Kadanoff, 1999); but compared with exponential distribution, the power-law distribution implies complexity (Barabasi, 2002; Barabasi and Bonabeau, 2003). Moreover, fractal dimension is a complexity measure, which can be related mathematically to degree of complexity (Ryabko, 1986). The urban density of Hangzhou seems to come between simple state and complex state at present (Figure 6). In the future, it may evolve into complex state through self-organizing process (see Appendix 3).

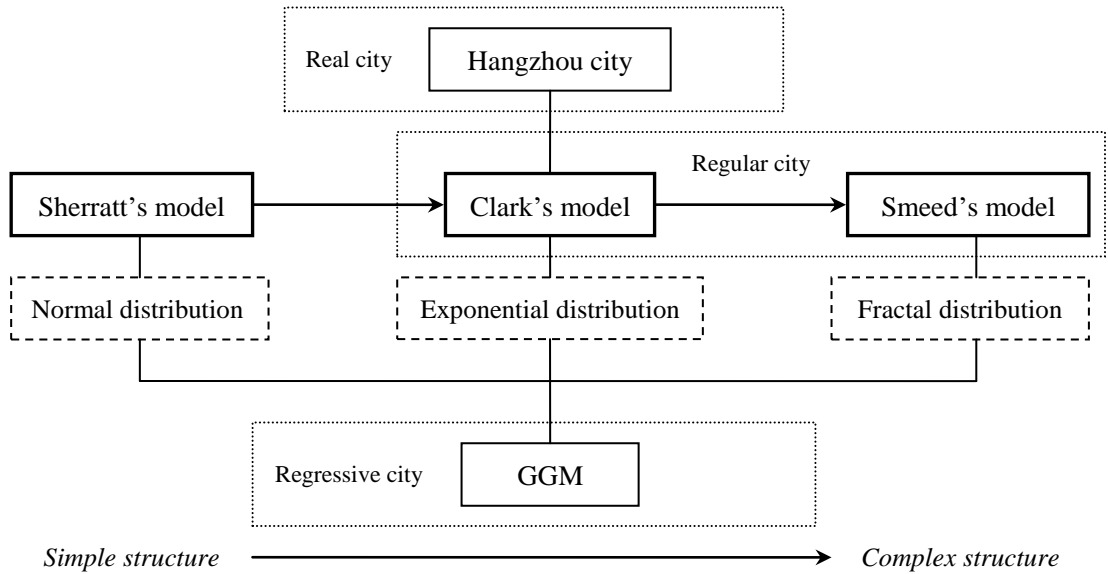

**Figure 6 A sketch map of urban evolution of population density from simple to complex structure**

## 4.3 Fractal, complexity, and sustainable cities

In many countries, especially developing countries, urban man-land relations became more and more strained in large cities due to population explosion. For example, fast urbanization results in deficiency of usable land and freshwater in numerous cities of China. Urban population suffered exponential growth while available land and water resources are limited. How to make reasonable use of urban land and fresh water is of vital importance to urban sustainability. The ideas from fractals provide a possible or potential way out. First, fractal bodies can occupy space in the optimum mode. If a city is designed or planned by means of fractal concepts, it will possibly fill up the geographical space in the best way so that urban land is economized for future development (Chen, 2008a). Second, both cities and rivers enjoy similar fractal scaling relations. Fractal theory can be used to harmonize the cities with natural environments including rivers (Chen, 2009). Thus the fresh water resources can be distributed and utilized in reason in urban and rural regions. Third, as indicated above, fractal organization can be employed to reconcile the equity of individuals with efficiency of the whole of cities.

Central place systems and rank-size distribution of cities are demonstrated to be fractals (e.g. Arlinghaus, 1985; Batty and Longley, 1994; Frankhauser, 1990). Potentially, the central place fractal and Zipf's law can be used to optimize intercity relations in a region, i.e., hierarchy and network of cities. The Smeed's model is associated with self-similar fractal form, while the Clark model maybe indicates some self-affine fractals. Both these model can be potentially used to optimize intra-urban structure. Fractal methods can be applied to both monocentric cities and polycentric cities. The fractal form of polycentric cities can be measured with the grid method, while the fractal growth and form of monocentric cities can be researched with radial method or grid method (Frankhauser, 1998). The Clark's model and Smeed's model are based on radial method. So this paper is only involved with monocentric cities.

For the developed countries with lower population density, the Clark's model or Smeed's model can be employed to optimize the urban population distribution. In contrast, for the developing countries with higher population density, we should use Smeed's model to optimize urban form in more complex way. The Clark's model of urban density often suggests some kinds of hidden fractal structure (Chen, 2008b). Compared with the negative exponential distribution, the inverse power law distribution of population takes on clear fractal pattern and thus can make the best of geographical space. Batty and Kim (1992) presented a significant idea, "form follows function". They argued that urban population density should be modeled with inverse power law rather than negative exponential law. My argument is that real cities may follow the power law or exponential law, but the power law can be employed to design future regular cities for populous regions (Figure 7). In theory, we can make the use of city planning and self-organized process to urge urban form to evolve into the inverse power law distribution.

Fractal optimization of urban form and structure is based on two keys, one the self-similar distribution, the other is appropriate fractional dimension. If the urban form satisfies the fractal distribution, the value of the fractal dimension will play an important role in urban function and efficiency. Suppose that a city suffers no influence from physical phenomena such as mountain and sea. If the fractal dimension of the city is too low, say, $D_f$<1.5, the urban space will be underfilled and urban land use will be wasteful. On the contrary, if the fractal dimension is too high, say, $D_f$>1.9, the urban space will be overfilled and urban land will be deficient. To characterize urban spatial structure, we must evaluate the fractal dimension of urban form. The main purpose of the new gamma model is to estimate the latent dimension and predict the natural evolutive direction (from simple structure to complex structure or from complex structure to simple structure). As space is limited, the related questions will be discussed in future studies.

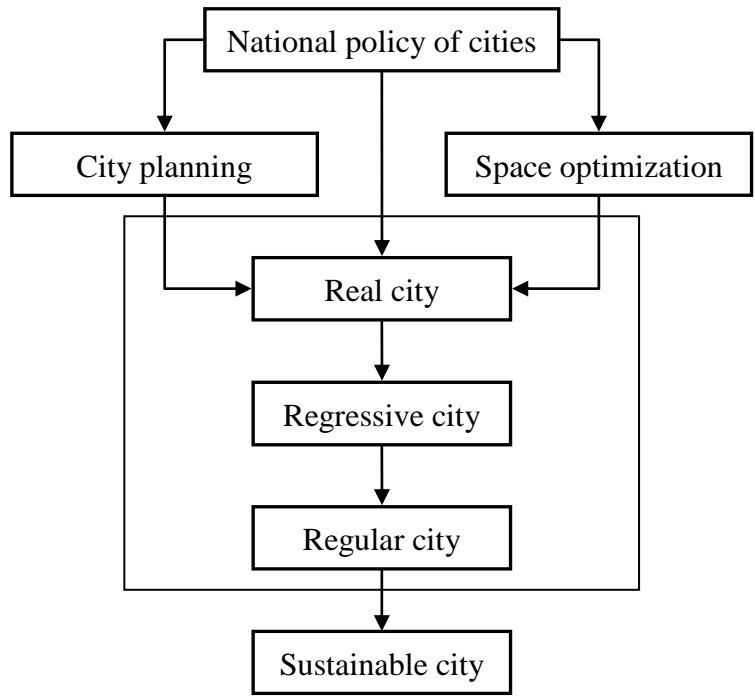

**Figure 7 National policy, city planning, and urban sustainable development**

# 5 Conclusions

Modeling urban density is not a new question, but many questions in this field are still pending further discussion. In fact, few human geographical problems have generated more research than the urban population density problem. However, a great deal of research is no guarantee of essential progress. This paper represents a new attempt to modeling urban population density using the ideas from fractal theory. The GGM has at least four uses: the first is to reconcile the debates between the negative exponential model, inverse power law, and normal model of urban density; the second is to characterize urban man-land relations; the third is to estimate the latent fractal dimension and the fourth is to predict spatial complication of urban evolvement. The main points of this research can be summarized as follows.

Firstly, it is not necessary for us to find the best model for urban density. Complex systems such as cities can be modeled with multiple mathematical equations. The diversity of models and variability of model parameters suggest complexity of city systems. Different models reflect different state of urban development and have different purposes in urban studies. The new model can be used to predict the spatial complication of urban evolution from simple to complex and then to more complex structure. The Gaussian distribution suggests a simple structure, the exponential distribution suggests a complex structure, and the power distribution suggests an even more complex structure. As to the GGM, when the constraint parameter ($\sigma$) varies from 2 to 1, the weighted Gaussian distribution will change into the negative exponential distribution; when the scale parameter ($r_0$) approaches infinity, the negative exponential distribution will evolve into the power-law distribution. The power law of cities is always associated with fractals and spatial complexity.

Secondly, we can use GGF to estimate the hidden fractal dimension of urban form. If the power-law distribution is concealed by quasi-exponential distribution, we can estimate the fractal parameter value by using the GGM. The fractal dimension is termed "latent dimension" in this context. Fractals suggest the optimized structure of city systems, and the fractal dimension is an indication of spatial optimization. If the fractal dimension of urban form is too high, the urban space will be overfilled and many urban problems such as traffic congestion and smog will become severe; in contrast, if the fractal dimension is too low, the urban space will be underfilled and the urban land use will be wasteful. The proper fractal dimension is one of the preconditions for spatial optimization of cities.

Thirdly, fractals and spatial complexity can play the basic role in the studies of urban sustainability. Cities as systems and systems of cities are complex spatial systems (Allen, 1997; Chen, 2008a; Wilson, 2000). Without the theory of spatial complexity, we could not comprehend urban systems, and thus could not solve many essential problems for sustainable development. On the other hand, the emergence of fractal patterns from urban form is a process of spatial complication. Fractal dimension is a measure of complexity, and fractal theory is a powerful tool for exploring spatial complexity. Especially, the theory of fractal cities has a potential application to urban optimization. The idea from fractals can be used to design better urban structure, improve urban man-land relations, reconcile the equity of individuals with efficiency of the whole of cities, and so on. Therefore, the studies of city fractals can lay a theoretical foundation for the studies of urban sustainable development.

**Acknowledgements**: This research was sponsored by the Natural Science Foundation of Beijing (Grant No. 8093033) and the National Natural Science Foundation of China (Grant No. 40771061). The support is gratefully acknowledged. The author would like to thank Dr. Jian Feng at Peking University for providing essential material on the urban density of Hangzhou. Many thanks to two anonymous referees whose comments have been very helpful in preparing the final version of this article.

# Appendices

## 1. The urban population density data of Hangzhou and negative exponential decay patterns

The urban population density data of Hangzhou in four years are processed by Feng (2002). These data are listed in Table A1. Fitting the four sets of data to the negative exponential model, the effect of point-line match is displayed in Figure A1.

**Table A1 Average population density and related variables of Hangzhou city in four years**

| Serial numbers | Radius ($km$) | Average density in 1964 | Average density in 1982 | Average density in 1990 | Average density in 2000 |
|---|---|---|---|---|---|
| 1 | 0.3 | 24130.876 | 29539.752 | 29927.903 | 28183.726 |
| 2 | 0.9 | 18965.755 | 22225.009 | 26634.162 | 26820.717 |
| 3 | 1.5 | 16281.905 | 18956.956 | 22261.980 | 24620.991 |
| 4 | 2.1 | 16006.650 | 19232.148 | 21611.817 | 23176.394 |
| 5 | 2.7 | 13052.016 | 15439.141 | 17290.295 | 18909.733 |
| 6 | 3.3 | 8259.322 | 9920.236 | 13178.503 | 19600.961 |
| 7 | 3.9 | 5798.447 | 7025.973 | 10537.808 | 16945.193 |
| 8 | 4.5 | 2625.945 | 3460.688 | 5559.761 | 10829.321 |
| 9 | 5.1 | 2142.703 | 2807.245 | 4180.368 | 7282.387 |
| 10 | 5.7 | 2141.647 | 2688.650 | 3923.003 | 6199.832 |
| 11 | 6.3 | 2185.160 | 2566.408 | 3515.837 | 5644.371 |
| 12 | 6.9 | 1438.027 | 1692.767 | 2197.220 | 4297.363 |
| 13 | 7.5 | 1083.473 | 1371.370 | 1795.763 | 3806.092 |
| 14 | 8.1 | 967.470 | 1256.167 | 1633.675 | 3152.766 |

| 15 | 8.7  | 842.494 | 1114.351 | 1442.105 | 2683.454 |
| 16 | 9.3  | 847.713 | 972.801  | 1265.412 | 2354.300 |
| 17 | 9.9  | 817.662 | 1050.963 | 1163.341 | 2028.299 |
| 18 | 10.5 | 812.050 | 1050.953 | 1143.197 | 1827.775 |
| 19 | 11.1 | 807.251 | 1050.998 | 1160.184 | 1651.076 |
| 20 | 11.7 | 625.112 | 979.407  | 1092.903 | 1580.848 |
| 21 | 12.3 | 691.323 | 901.339  | 1006.045 | 1490.260 |
| 22 | 12.9 | 574.569 | 870.020  | 972.123  | 1465.000 |
| 23 | 13.5 | 532.355 | 665.846  | 816.501  | 1278.000 |
| 24 | 14.1 | 381.306 | 486.856  | 679.057  | 1033.000 |
| 25 | 14.7 | 369.036 | 489.208  | 581.566  | 958.000  |
| 26 | 15.3 | 375.204 | 456.473  | 563.203  | 882.000  |

**Data source**: From Feng (2002).

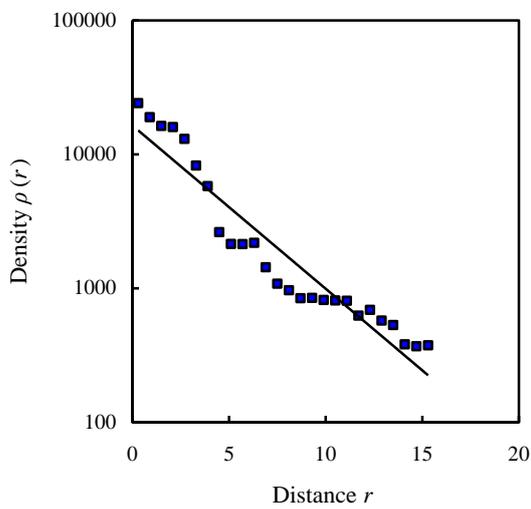

a. 1964

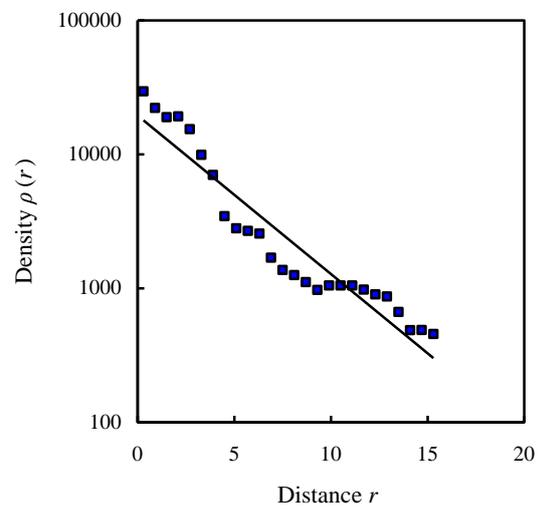

b. 1982

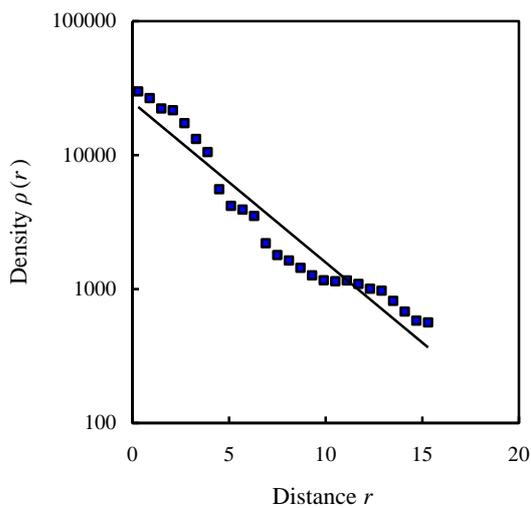

c. 1990

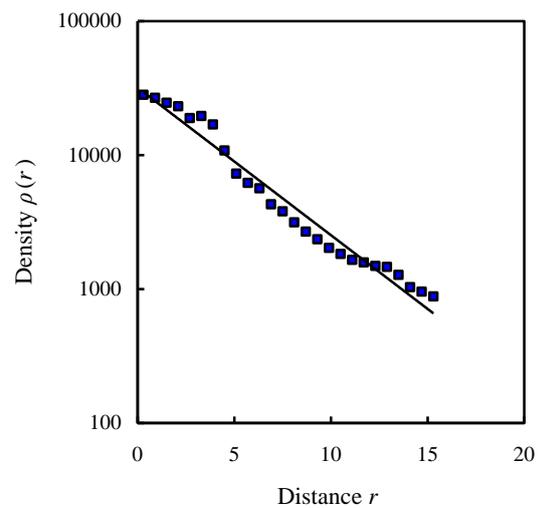

d. 2000

**Figure A1 The semi-logarithmic plot of population density of Hangzhou in four years**

## 2. Which model is more suitable for the urban density of Hangzhou city?

The choice of model can be decided by comparing adjusted correlation coefficient square and the standard error of three models: the negative exponential model, the exponential-power model, and the weighted exponential-power model (GGM). As far as the adjusted $R$ square and standard error are concerned, the exponential-power model is the best one, and the negative exponential model is not very good. GGM comes between the negative exponential model and the exponential-power model according to adjusted $R^2$ and standard error (Table A2). Where prediction is concerned, the exponential-power model is more advisable; but where explanation is concerned, the weighted exponential-power model is the best one.

**Table A2 The $R$ square, adjusted $R$ square, and standard error values of three urban density models in four years**

| Models | Statistics | 1964 | 1982 | 1990 | 2000 |
|---|---|---|---|---|---|
| Negative exponential model (NEM) | $R^2$ | 0.907 | 0.904 | 0.930 | 0.968 |
|  | Adjusted $R^2$ | 0.903 | 0.900 | 0.927 | 0.967 |
|  | Standard error | 0.421 | 0.416 | 0.353 | 0.215 |
| Exponential-power model (EPM) | $R^2$ | 0.958 | 0.957 | 0.965 | 0.976 |
|  | Adjusted $R^2$ | 0.956 | 0.955 | 0.963 | 0.975 |
|  | Standard error | 0.283 | 0.280 | 0.252 | 0.188 |
| New model (GGM) | $R^2$ | 0.945 | 0.944 | 0.953 | 0.971 |
|  | Adjusted $R^2$ | 0.941 | 0.940 | 0.949 | 0.968 |
|  | Standard error | 0.329 | 0.323 | 0.297 | 0.211 |

**Note**: Owing to different degree of freedom, the goodness of fit of different models is not comparable in general. However, the adjusted correlation coefficient square is comparable because that the degree of freedom has been "punished" in the formula.

## 3. More empirical evidences for the GGM of urban density

The GGMs and the related method of fractal dimension estimation for urban form can be applied to other cities, especially the concentric cities, in the world. For example, we can fit the population density data of Beijing city, China, in 1982 and 1990, to equation (8). The fractal dimension of land-use patterns are estimated and tabulated as follows (Table A3). The results also suggest that the urban land use expanded as both the fractal dimension and the characteristic radius grew. It can be seen from the scatter plot that the data points are distributed along a trend line approximately, except for the first point indicative of the "central political district (CPD)" (Figure A2). CPD bears an analogy with the central business district (CBD) in the Western cities, where the resident population is always smaller than what is expected (Clark 1951). Therefore, the first data point as an outlier is usually rejected (Clark 1951; Banks 1994). In fact, Beijing is not a typical concentric city, and the modeling result is not good enough to be taken as an example. This lends further support to the judgment that the general gamma model should be applied to the monocentric cities.

**Table A3 Estimated values of model parameters for urban density of Beijing city, 1982 and 1990**

| Parameter and statistic | 1982 | 1990 |
|---|---|---|
| Proportionality constant $C$ | 55482.596 | 43270.709 |
| Characteristic radius $r_0$ | 8.163 | 8.794 |
| Constraint parameter $\sigma$ | 1.200 | 1.450 |
| Fractal dimension $D_f$ | 1.440 | 1.555 |
| $P$-value (significance) | 0.046 | 0.039 |
| Determination coefficient $R^2$ | 0.936 | 0.950 |

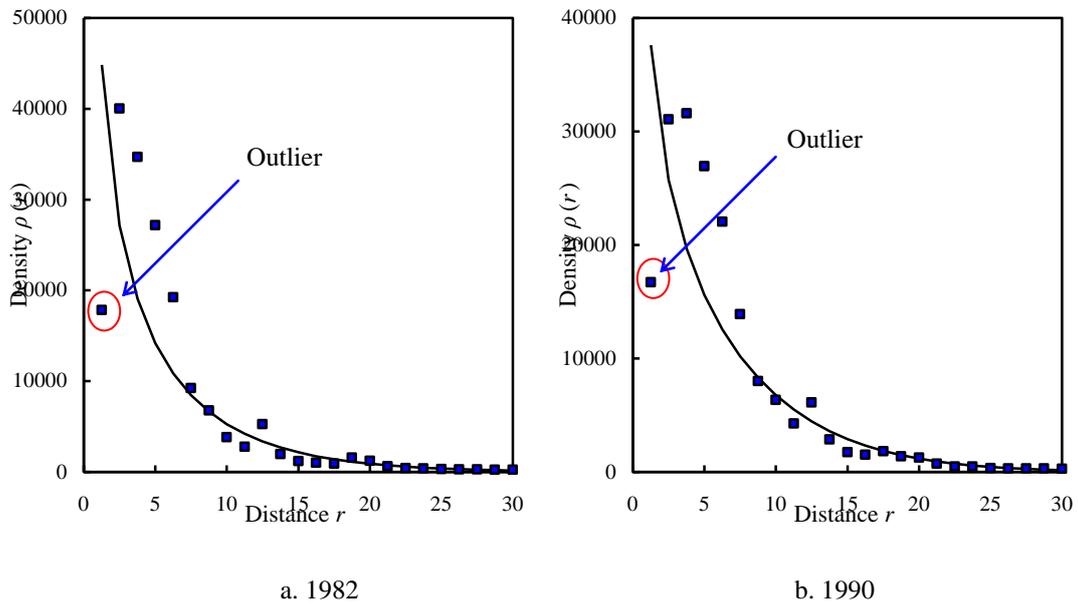

a. 1982    b. 1990

**Figure A2 The arithmetic plots based on the GGM of population density for Beijing, 1982 and 1990**

**Note:** Like the Clark's model, the "GGM" is made for the population density of monocentric cities. Beijing is a quasi-polycentric city with multiple nuclei, thus, the modeling effect is not satisfactory.